# Opto-Mechanical Design of ShaneAO: the Adaptive Optics System for the 3-meter Shane Telescope

C. Ratliff , J. Cabak, D. Gavel, R. Kupke, D. Dillon, E. Gates, W. Deich, J. Ward, D. Cowley, T. Pfister, M. Saylor

UCO Lick Observatories, University of California, Santa Cruz, 1156 High St. Santa Cruz, CA 95064 USA

## ABSTRACT

A Cassegrain mounted adaptive optics instrument presents unique challenges for opto-mechanical design. The flexure and temperature tolerances for stability are tighter than those of seeing limited instruments. This criteria requires particular attention to material properties and mounting techniques. This paper addresses the mechanical designs developed to meet the optical functional requirements. One of the key considerations was to have gravitational deformations, which vary with telescope orientation, stay within the optical error budget, or ensure that we can compensate with a steering mirror by maintaining predictable elastic behavior. Here we look at several cases where deformation is predicted with finite element analysis and Hertzian deformation analysis and also tested. Techniques used to address thermal deformation compensation without the use of low CTE materials will also be discussed.

## I. INTRODUCTION

When a new adaptive optics system for the Lick three-meter telescope was designed and fabricated, a main objective was to improve angular resolution and work with shorter light wavelengths. The key upgrades included:

1. High order wavefront correction using Boston Micromachines 32x32 actuator MEMs deformable mirror;

2. High resolution infrared detector: Teledyne Hawaii 2RG (2K x 2K);

3. New fiber laser for producing the reference guide star;

4. Stiff mechanical support structure with stiff and precise motion stages to allow for easier acquisition and longer exposures.

This paper addresses item 4, improvement to the support structure and motion stages. Other UC Observatory papers in these proceedings address items 1, 2, and 3 (see references 1 and 2.)

The first generation Lick 3-meter adaptive optics system was optimized for astronomical imaging and spectroscopy in K-band (2.2 microns). Achieving diffraction limited results at shorter wavelengths (J-band and H-band) requires higher-spatial frequency sampling of the wavefront and a larger number of degrees of freedom on the deformable mirror[5]. Shorter wavelengths also produce higher angular resolution at the diffraction limit ($\lambda/D$). Both of these requirements dictate tighter tolerances in stability at both pupil and image conjugates. Images will not reach the diffraction-limit if there is excessive flexure/deformation during the exposure, especially if there is differential flexure between the wavefront sensor, deformable mirror, and/or science imaging device.

The effects of flexure become especially problematic during long camera exposures. All structural systems experience some deformation with a change in applied load and a Cassegrain mounted telescope instrument will experience a continual change in gravitational load direction while tracking a celestial object.

Figure 1 shows the Shane telescope fork axis (dashed line), which the telescope and instruments rotate about as a celestial object is tracked during an exposure. Since the Cassegrain instrument mount, located under the primary mirror, also rotates, there are two axes of motion that affect gravity induced flexure in the AO system. For most of the AO stages and mounts, we designed for a worst-case change in gravitational load which would correspond to a 3 hour camera exposure; a change in gravity load by roughly 45 degrees.

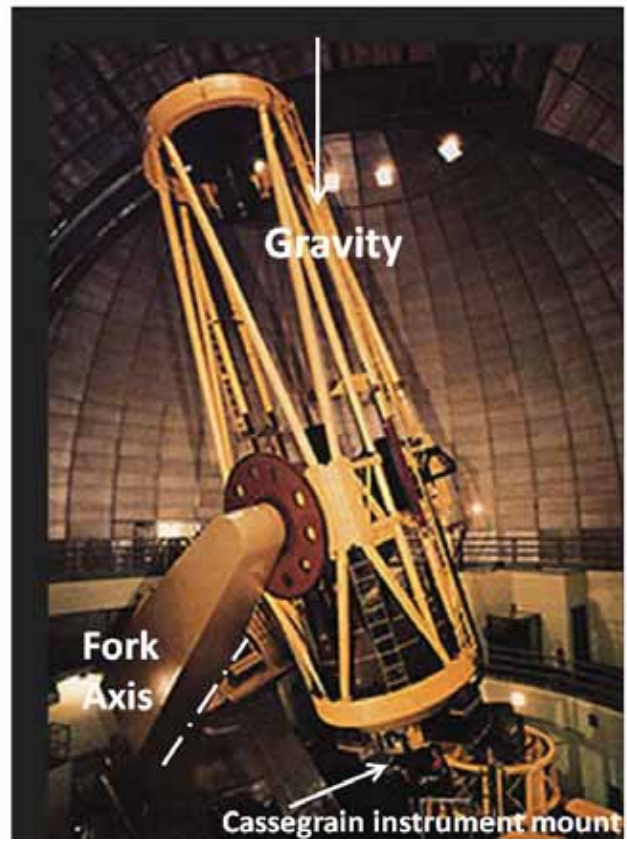

**Figure 1: Gravity Vector and Fork Mounted Rotation Axis**

## II. DESIGN APPROACH

Several designs will be presented to show our approach to designing positioning stages and structural support systems to meet our functional requirements. These designs include:

1. A 5 axis stage for the 32x32 MEMs deformable mirror, required to align the deformable mirror with the wavefront sensor;
2. Custom designed x-y flexure stage for the wave front sensor;
3. AO bench support structure.

The location for the MEMs and wavefront sensor are shown in Figure 2, the center ray of the telescope light is represented by the light blue line.

The flexure error budget for each subassembly is outlined in Table 1.

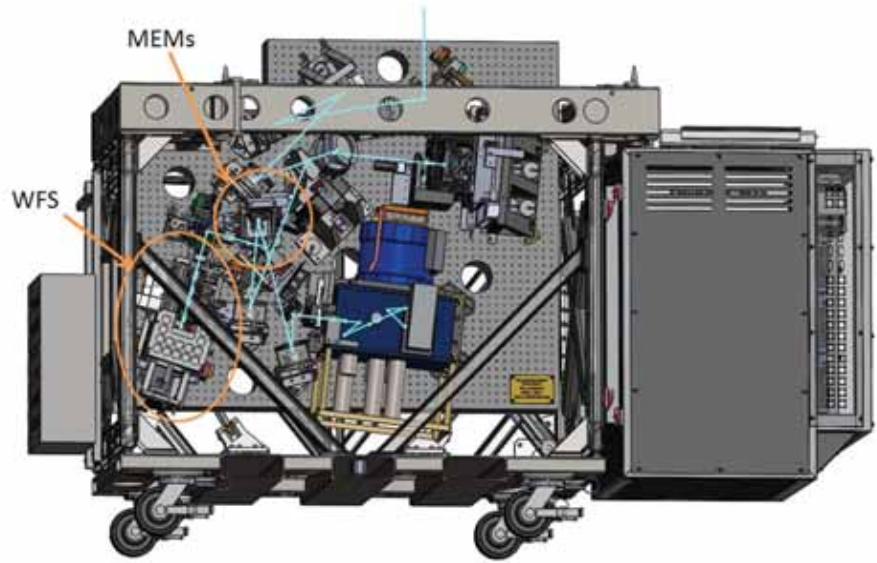

**Figure 2: MEMs and WFS in AO System**

**Table 1: Opto-Mechanical Stability Requirements**

| Name | Description | DOF | Stability |
|---|---|---|---|
| MEMs | "Tweeter" MEMs high order deformable mirror with 32x32 actuators. | 5 | 2" over 3 hrs |
| WFS Dichroic | Splits light into wavefront sensor (WFS). Dichroic changer will select one of three dichroics depending on science wavelength and LGS/NGS mode. | 1 | 4" over 1/2 hour |
| WFS -CAM X | X translation of WFS Camera | 1 | 0.1µm linear stability |
| WFS -CAM Y | X translation of WFS Camera | 1 | 0.1µm linear stability |
| WFS -Lenslet Selector | X translation of WFS Camera | 1 | 0.1µm linear stability |
| WFS -Focus Z | X translation of WFS Camera | 1 | 1-2 µm |

## 1. MEMS STAGE

The MEMS 5 axis positioning stage design is based on a Newport 9082 5 axis aligner mounted on a custom aluminum base. It consists of a plate supported by 6 spheres and springs that load the plate against the spheres. The position of each sphere is controlled by a fine pitch screw/wedge/motor mechanism that controls the position of each sphere in a direction that is normal (or nearly so) to the plate. This stage was chosen given the small footprint allowance in the optical prescription and large number of degrees of freedom to control.

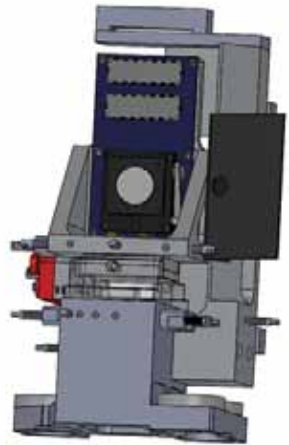

Figure 3: MEMs Alignment Stage

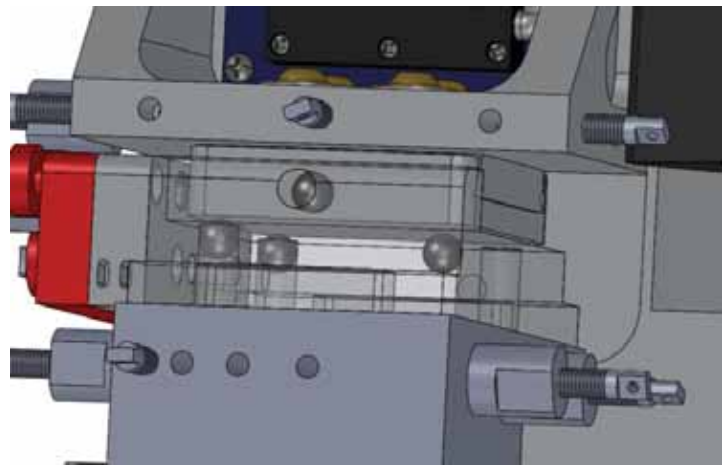

Figure 4: Close Up View of Kinematic Translation Stage: Wedges Omitted, Base Plate is Shown as Transparent

There was a problem with the off the shelf design because it allowed too much deformation under varying gravity loads, measured to be 5x higher than requirements. We decided to try to improve the stock design by making it stiffer. The first thing to investigate was more preload on the kinematic interfaces. Based on some deformation analysis, it was estimated the spring preloading force would need to be roughly 5x higher.  With this higher preload, there would be more stress on the material.  If this stress were to cause plastic deformation/brinneling, positioning issues and possibly hysteresis in the positioning mechanism would result.  We therefore decided to investigate replacing the stock 6061 T6 aluminum material with several candidate materials which included M2 tool steel and sapphire pucks.  Figure 5 shows the stress in the aluminum for a 30 N load and 9.5mm diameter spheres.   The stress exceeds the 6061 T6 aluminum yield stress which is 40,000 psi (276 MPa).

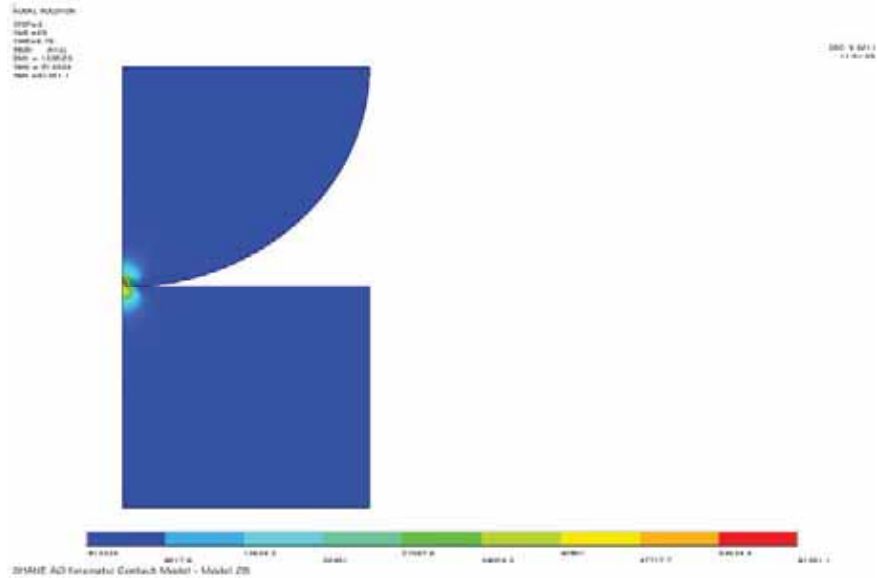

**Figure 5.  Aluminum: 30 N Spring Preload Yields Material: Stress Exceeds 40,000 psi**

M2 tool steel turned out to have some margin of safety with respect to yielding for the same load and although less stiff than sapphire, it is much less expensive and quicker to fabricate.  After fabrication we tested our design with the setup shown in  Figure 6.  It consists of an alt-azimuth index table, linear variable differential transformer displacement transducer with 0.1 micron resolution and the modified stage.

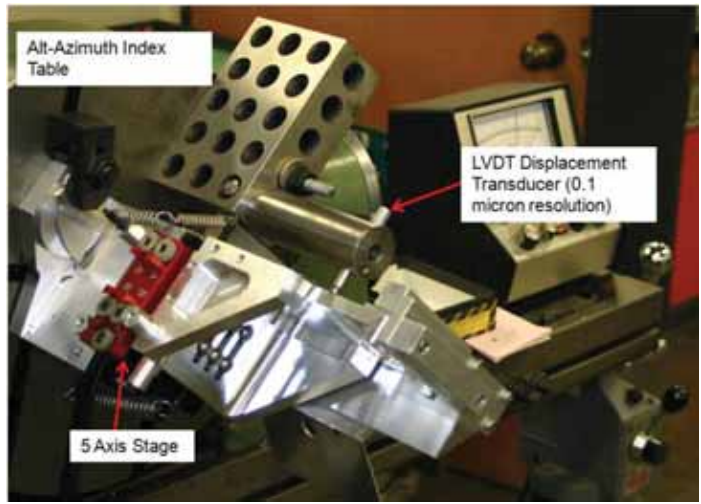

**Figure 6. Flexure Measurement**

Table 2: Measured Deformation Results

|  | Case1<br>Stock Kinematic Stage | Case2<br>M2 Tool Steel and 6 to 11 lb preload | Case3<br>M2 Tool Steel and 15 to 20 lb preload |
|---|---|---|---|
| **Maximum Deflection (μm) within 22.5° of Zenith** | 3.7 | 1.4 | 0.8 |
| **Maximum Angular Deflection (arcsec) within 22.5° of Zenith** | 7.3 | 2.8 | 1.6 |
| **Maximum Deflection (μm) within 45° of Zenith** | 4.6 | 2.2 | 1.5 |
| **Angular (arcsec) Hysteresis** | 5.9 | 2.2 | 0.5 |

In summary, we were able to transform a stock 5-axis kinematic stage into a much stiffer one. M2 tool steel was selected based on its cost, high modulus of elasticity, high strength, and ease of machining. Both the plate supported by spheres and translating wedges were replaced with this high strength steel. All mating surfaces were machined to a surface finish of approximately 10 microinches RMS. Preload spring forces were increased by a factor of 3-6 depending on orientation. The angular flexure was reduced to less than 10% of the stock stage. Experiments showed a much stiffer stage with no plastic deformation. The original stage with the stock aluminum and brass components exhibited less stiffness and considerable plastic deformation.

## 2. WAVEFRONT SENSOR FLEXURE STAGE

The Shane adaptive optics system utilizes a Shack-Hartmann wavefront sensor with selectable configurations. Depending on light levels, the user may choose an 8x8 or 16x16 array across-the-pupil subaperture arrangement, with an upgrade path to 32 subapertures across available in the future.

The wavefront sensor consists of a fast 160x160, 21 μm pixel Lincoln Laboratories CCD detector in a SciMeasure camera head. To control fine positioning on the sub pixel level, we mounted it on a custom x-y flexure stage. In front of the camera is the selector stage that allows choosing between optics that provide 8x or 16x Hartmann lenslet sampling across the beam. There is an open slot that may be occupied by a 32x lenslet barrel in the future. The x-y flexure stage on the camera head allows fine alignment to the lenslet barrel selected.

The AO system will be used with both natural guidestars (NGS) and laser guidestars (LGS), therefore a requirement is that the entire wavefront sensor (lenslet optics and camera) ride on a focus stage to accommodate the shift in focus between NGS and LGS. Each item is indicated in Figure 7.

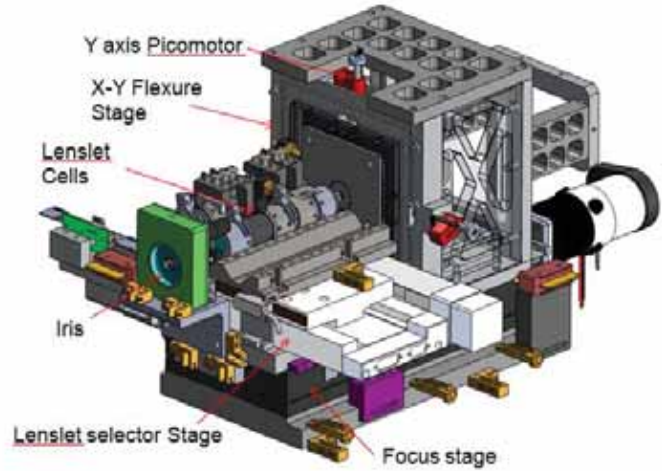

**Figure 7: Wave Front Sensor**

A fairly demanding stability requirement, 0.1 micron linear deformation for a 30 minute exposure, for this wave front sensor requires attention to certain aspects of the design. We decided on a custom flexure stage since space was limited and it needed to mate with the Scimeasure CCD camera head mount geometry. Flexure stages can provide high precision positioning with a high stiffness in axes other than the direction of travel. For driving the x-y flexure stage we chose an off the shelf product from Newport, a Picomotor™, which provides open loop control with 30 nm resolution. The two other stages, the lenslet selector and focus stages, provide closed loop control with sub 100nm resolution and sub 2 micron resolution.

One thing that had to be considered in this design was the Hertzian deformation of the Picomotor spherical tip against the flexure stage contact area. We found out that the spherical tip pushing against a flat contact would dominate the overall deformation for a 7.5 degree change in the gravity vector relative to the reference frame of the flexure stage.

We were able to dramatically reduce the Hertzian deformation by having the Picomotor tip interface with a spherical receiving socket. A comparison of the Hertzian deformation for several cases is shown in Table 3. A load change of 1.7 N corresponds to a worst case change in gravity load by 7.5 degrees. A section view of the final design is shown in Figure 8. This improvement reduced the deformation by 95%. Worth noting is the Picomotor was near its axial force limit with the amount of preload used. In hindsight an actuator with substantially more axial force capability than the required preload would have been preferred.

**Table 3: Comparison of Change in Deformation for Load Change of 1.7 N**

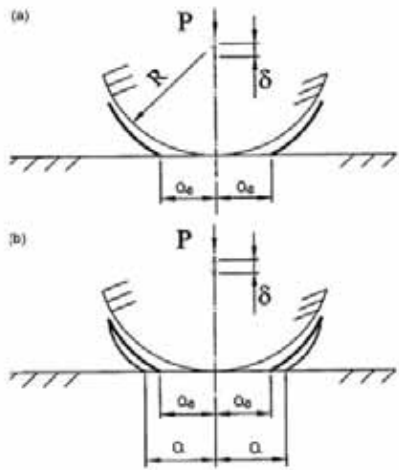

| Summary of Results: $\Delta P=1.7$ Newton | Change in deflection=$\Delta\delta$ Allowance=0.1microns |
|---|---|
| Picomotor on Aluminum Flat | 2 microns |
| Picomotor on Tool Steel | 1.7 microns |
| Picomotor in 5mm dia. Spherical Socket | .05 microns |

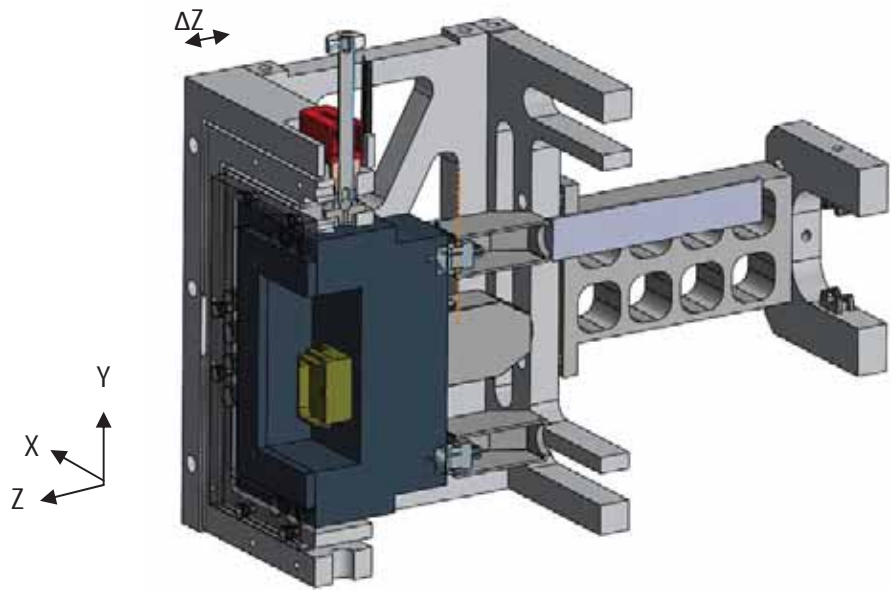

**Figure 8: Section View Camera Flexure Stage Design**

Another important aspect in designing this flexure stage is determining thicknesses. Our error budget in Z (as shown in Figure 8) is 1 micron, 10x that in the x and y, but still an important requirement to meet. Finite element analysis was used to determine if the as-designed thickness, labeled ΔZ in Figure 8, is appropriate. Again the change in load was a 7.5 degree change in the gravity vector, rotated about the X axis. The change in the Z deformation for this case was small, about .04 microns, and well within specifications. As it turned out, the gravity load changes were not the most important factor in designing this aspect of the flexure stage. External cable forces, which are highly dependent on stress relief implementation, drove the decision to keep the ΔZ value the same as that shown in the FEA rather than make the flexure stage thinner.

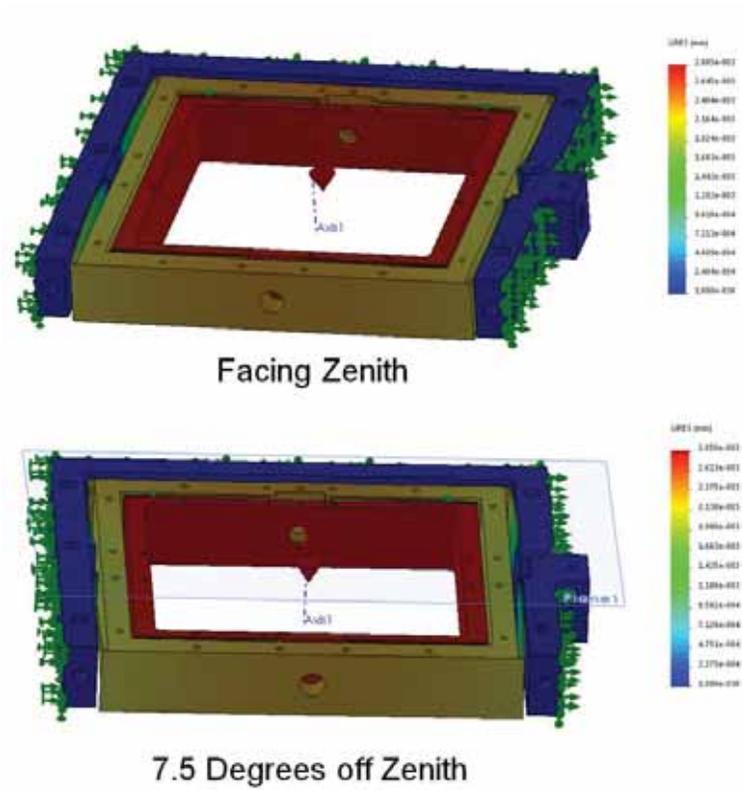

**Figure 9: FEA Results, 7.5 Degree Change in Gravity Vector**

## 3. SUPPORT STRUCTURE

The support structure consists of several systems: the AO opto-mechanical support assemblies, the optical table, and the optical table support system. The AO opto-mechanical support assemblies include anything that interfaces with the optical beam or calibration of the beam. It consists of many subassemblies, two of which are discussed above. In addition to the deformable mirrors the assemblies include 30 axes of precision linear and rotary motion.

The optical table is a custom sandwich structure that consists of 400 series steel plates bonded and welded to a trussed steel honeycomb interior. The table is all steel rather than a low CTE material like Invar. This table is the base support for all of the AO opto-mechanical support assemblies.

The optical table support structure is shown in Figures 10 and 11. Six struts support the table that anchor into several welded and bolted steel frames that interface with the telescope. Figure 10 shows the optical path thru the system. Figure 11 shows the six struts connecting to the rear surface of the optical table.

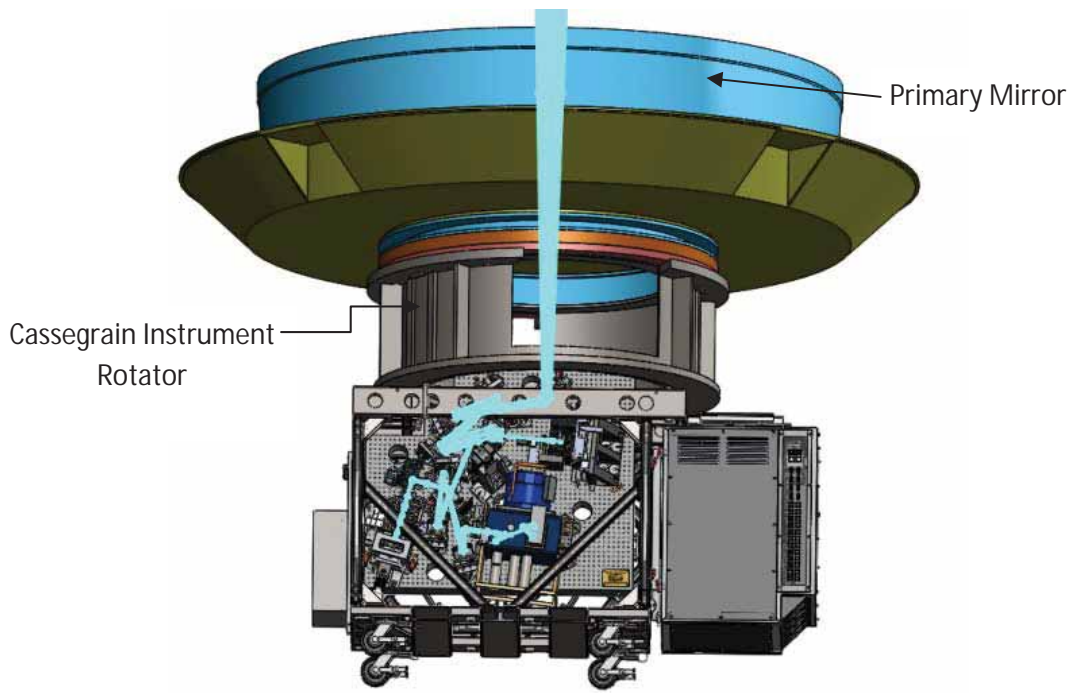

**Figure 10: Front Side of AO System, Telescope Light Highlighted Blue**

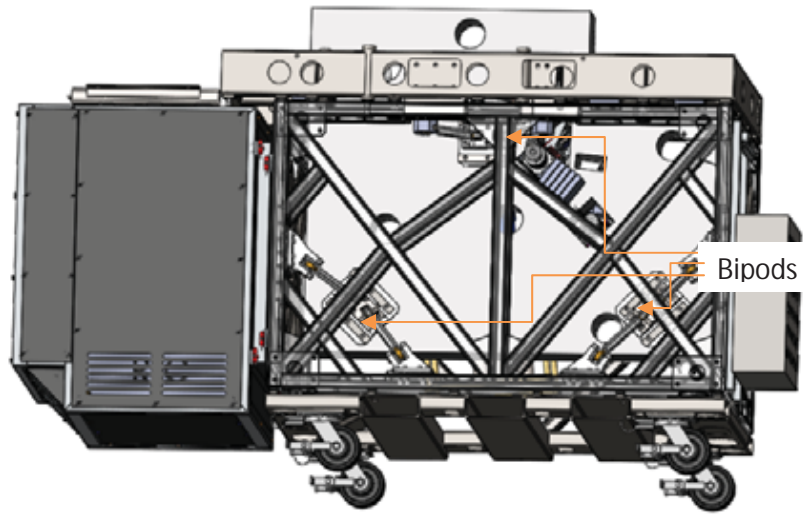

**Figure 11: Backside View**

The six support struts form three bipods. One of these bipods is shown in Figure 12. The axes of the bipod struts intersect at the mounting surface. The concept of intersecting at the surface with spherical bearing support at the end of each strut eliminates transmission of moment loads to the optical table. This design reduces both bending stress imposed on the table and flexure which could cause misalignment between optical components. The goal is to support the table as a simple 6-degree-of-freedom rigid body and avoid introducing moments, which could add unwanted bench deformation.

Each strut that makes up the bipods is adjustable in length along the strut axis with a fine thread turnbuckle design. This adjustability allows for kinematic positioning of the optical table in all 6 degrees of freedom and allowed for relatively easy alignment of the first steering mirror to the center ray of the telescope.

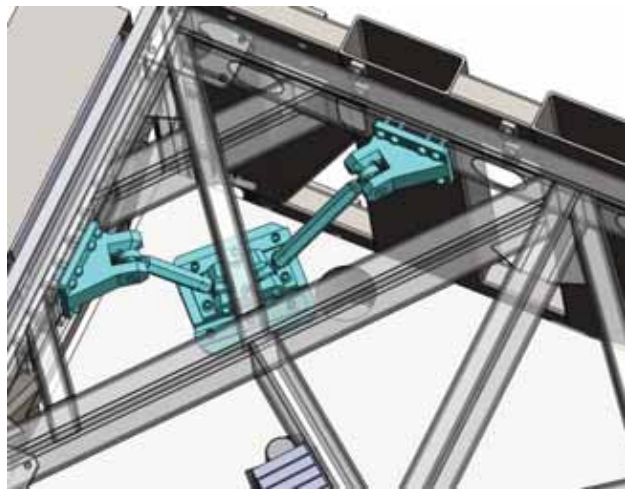

**Figure 12: Optical Bench Support Bipod, Highlighted in Blue**

Another aspect of the design was to ensure we maintained elastic deformation for all load conditions. One such scenario was earthquake loading. We looked at the response spectrum to a 2 G load with a forcing function between 0.5Hz and 100 Hz. The stress results (psi) are shown in Figure 13 for one direction of load excitation. The resonant frequencies for all of the significant modes are shown in Table 4. The results show that the stress is less than 10,000 psi for key structural components.

**Table 4: Resonant Vibration Frequencies Under 100 Hz**

| Mode | Frequency (Hz) |
|------|----------------|
| 1    | 26.643         |
| 2    | 29.256         |
| 3    | 30.698         |
| 4    | 46.885         |
| 5    | 53.978         |
| 6    | 63.328         |
| 7    | 78.720         |

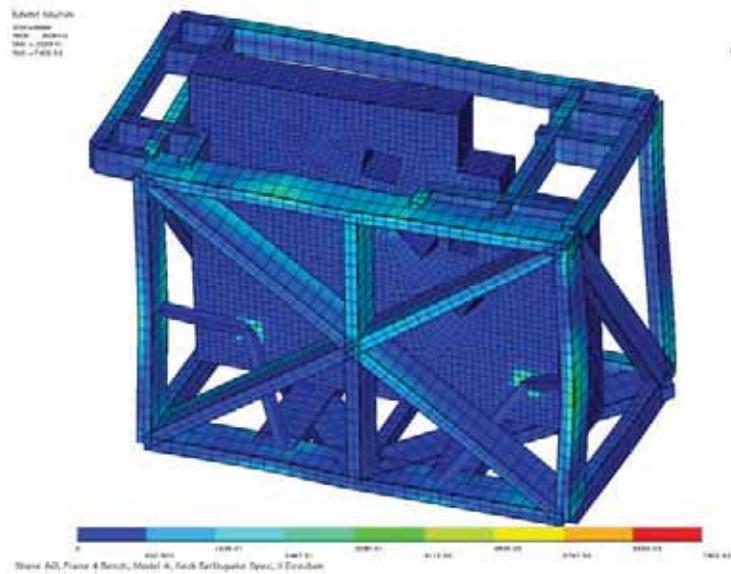

**Figure 13: Resulting Stress for Vibration Mode of Maximum Displacement (in psi)**

Elastic deformation ensures there is no need to realign after exposure to these types of loads, and that any residual flexure motion is repeatable. This allows the AO system to use active compensation using steering mirrors to account for drift in the optical axis as the gravity vector changes.

The thermal behavior of the support structure is no less important than the gravitational flexure. In our design, one objective was to ensure that we are insensitive to typical temperature swings that occur at the observatory while minimizing the use of more costly low coefficient of thermal expansion materials like Invar and glass ceramics. Again, the thermal behavior is designed to be repeatable so that the AO system can actively compensate beam wander as a function of temperature.

The material coefficients of thermal expansion (CTE) for the welded frame (1020 steel), struts (stainless steel), optical bench (carbon steel and 400 stainless), and first steering mirror (stainless steel) were all selected to be close to that of the telescope itself (1020 steel). As a result, the system expands and contracts together. For a 15 C change in temperature, the struts supporting the table change length by over 60 microns. However the table and first fold mirror move less than 5 microns relative to the center beam of the telescope.

We also used the strategy of matched CTE's for corresponding input sensors and actuators for the AO assemblies. The stages and support structure of the wave front sensor and MEMs have matched CTE's. Both are made primarily from 6061 T6 Aluminum. Although they move more than other parts of the system that are made of steel, Invar, or Zerodur, they move together to maintain a 1:1 relationship between each wave front sensor quad cell and MEMs actuator group. The same is true of the tip-tilt sensor and tip-tilt deformable mirror. Using aluminum allows lower cost mounts, better selection of commercial translation stages, and less weight.

## III. CONCLUSION:

We have touched on some of the many design aspects to consider when building the opto-mechanical structure for a high precision astronomical instrument. AO systems improve angular resolution over seeing-limited instruments and this maps directly to putting more demanding precision and stiffness requirements on the opto-mechanical mounts and stages. The key to a successful design is to meet the functional requirements, which include motion range and precision, ease of use, and spatial constraints, while at the same time addressing the alignment problems introduced by load variations, vibration, or changes in temperature.

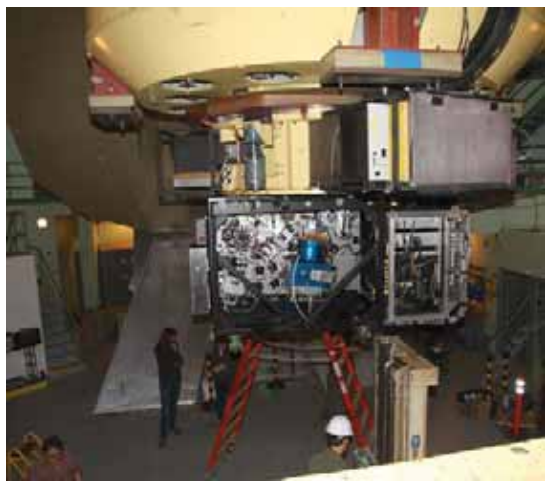

Figure 14: Shane AO on Telescope, Covers Removed